\documentclass{elsart}

\usepackage{graphicx}
\usepackage{subfigure}
\usepackage{amsmath}
\usepackage{xspace}

\def\NIM{Nucl. Instr. Meth.}

\newcommand{\BW}{\mathrm{BW}}
\newcommand{\GS}{\BW^{\mathrm{GS}}}
\newcommand{\mrho}{\mathrm{M}_\rho}
\newcommand{\momg}{\mathrm{M}_\omega}
\newcommand{\eepp}{\ensuremath{e^+e^-\to\pi^+\pi^-}\xspace}
\newcommand{\eeee}{\ensuremath{e^+e^-\to e^+e^-}\xspace}
\newcommand{\eemm}{\ensuremath{e^+e^-\to\mu^+\mu^-}\xspace}
\newcommand{\ee}{\ensuremath{e^+e^-}}

\begin{document}

\begin{frontmatter}

\title{Measurement of $e^+e^-\rightarrow \pi^+\pi^-$ cross
section with CMD-2 around $\rho$-meson}

\author[BINP]{R.R.Akhmetshin},
\author[BINP]{E.V.Anashkin},
\author[JINR]{A.B.Arbuzov},
\author[BINP,NGU]{V.M.Aulchenko},
\author[BINP]{V.Sh.Banzarov},
\author[BINP]{L.M.Barkov},
\author[BINP]{S.E.Baru},
\author[BINP]{N.S.Bashtovoy},
\author[BINP,NGU]{A.E.Bondar},
\author[BINP]{D.V.Bondarev},
\author[BINP]{A.V.Bragin},
\author[BINP]{D.V.Chernyak},
\author[YALE]{S.Dhawan},
\author[BINP,NGU]{S.I.Eidelman},
\author[BINP,NGU]{G.V.Fedotovich},
\author[BINP]{N.I.Gabyshev},
\author[BINP,NGU]{D.A.Gorbachev},
\author[BINP]{A.A.Grebenuk},
\author[BINP,NGU]{D.N.Grigoriev},
\author[YALE]{V.W.Hughes},
\author[BINP,NGU]{F.V.Ignatov},
\author[BINP]{S.V.Karpov},
\author[BINP]{V.F.Kazanin},
\author[BINP,NGU]{B.I.Khazin},
\author[BINP,NGU]{I.A.Koop},
\author[BINP]{P.P.Krokovny},
\author[JINR]{E.A.Kuraev},
\author[BINP]{L.M.Kurdadze},
\author[BINP,NGU]{A.S.Kuzmin},
\author[BINP,BU]{I.B.Logashenko},
\author[BINP]{P.A.Lukin},
\author[BINP]{A.P.Lysenko},
\author[BINP]{K.Yu.Mikhailov},
\author[BU]{J.P.Miller},
\author[BINP,NGU]{A.I.Milstein},
\author[BINP]{I.N.Nesterenko},
\author[BINP]{V.S.Okhapkin},
\author[BINP]{A.A.Polunin},
\author[BINP]{A.S.Popov}, 
\author[BINP]{T.A.Purlatz},
\author[BU]{B.L.Roberts},
\author[BINP]{N.I.Root},
\author[BINP]{A.A.Ruban},
\author[BINP]{N.M.Ryskulov},
\author[BINP]{A.G.Shamov},
\author[BINP]{Yu.M.Shatunov},
\author[BINP,NGU]{B.A.Shwartz},
\author[BINP]{A.L.Sibidanov},
\author[BINP]{V.A.Sidorov},
\author[BINP]{A.N.Skrinsky},
\author[BINP]{V.P.Smakhtin},
\author[BINP]{I.G.Snopkov},
\author[BINP,NGU]{E.P.Solodov},
\author[BINP]{P.Yu.Stepanov},
\author[BINP]{A.I.Sukhanov},
\author[PITT]{J.A.Thompson},
\author[BINP]{V.M.Titov},
\author[BINP]{Yu.Y.Yudin},
\author[BINP]{S.G.Zverev}
\address[BINP]{Budker Institute of Nuclear Physics, 
Novosibirsk, 630090, Russia}
\address[JINR]{Joint Institute of Nuclear Research,
Dubna, 141980, Russia}
\address[NGU]{Novosibirsk State University, 
Novosibirsk, 630090, Russia}
\address[BU]{Boston University, Boston, MA 02215, USA}
\address[YALE]{Yale University, New Haven, CT 06511, USA}
\address[PITT]{University of Pittsburgh, Pittsburgh, PA 15260, USA}
\vspace{-3mm}
\begin{abstract}
The cross section of the process  
\(e^{+}e^{-}\rightarrow \pi ^{+}\pi ^{-} \) 
has been measured using about 114000 events collected by the
CMD-2 detector at the VEPP-2M $e^+e^-$ collider in the 
center-of-mass energy range from 0.61 to 0.96 GeV. Results
of the pion form factor determination with a 0.6\% systematic 
uncertainty are presented. The following values of the $\rho$- 
and $\omega$-meson parameters were found: $\mrho=(776.09\pm 0.81)$ MeV,
$\Gamma_\rho=(144.46\pm 1.55)$ MeV,
$\Gamma(\rho\rightarrow e^+e^-)=(6.86\pm 0.12)$ keV,
$Br(\omega\rightarrow\pi^+\pi^-) = ( 1.33\pm 0.25 ) \%$.
Implications for the hadronic contribution to the muon anomalous
magnetic moment are discussed.
\end{abstract}

\end{frontmatter}

\section{Introduction}

The cross-section of the process $e^+e^- \to \pi^+\pi^-$ is usually
expressed in terms of the pion electromagnetic form factor $F_\pi(s)$: 
\begin{equation}
 \sigma_{\eepp} = \frac{\pi\alpha^2}{3s}\beta^3_\pi \left| F_\pi(s)
\right|^2,
\label{ffdef}
\end{equation} 
where $s$ is the center-of-mass (c.m.) energy squared, $m_{\pi}$ is the 
pion mass and $\beta_\pi=\sqrt{1-4m_{\pi}^2/s}$ is the pion velocity
in the c.m. frame.

The pion form factor measurement is crucial for a number of physics
problems. Detailed experimental data in the time-like region allow
a determination of the parameters of the $\rho(770)$ meson and its 
radial excitations. Extrapolation of the energy dependence of the 
pion form factor to the point $s=0$ provides a value of the pion
electromagnetic radius.  

In the energy range below $\phi(1020)$ the process \eepp gives
the dominant contribution to the quantity $R(s)$ defined as 
\[
R=\sigma (e^{+}e^{-}\rightarrow hadrons)/
\sigma (e^{+}e^{-}\rightarrow \mu ^{+}\mu ^{-}).
\]
$R(s)$ is an important measurable quantity widely used for various 
QCD tests as well as calculations of the dispersion integrals. 
For such applications, at high energies $R(s)$ is usually
calculated within the perturbative QCD frame, while for the low energy 
range the direct measurement of the $\ee\to{hadrons}$ cross-section  
is necessary.

Particularly, knowledge of $R(s)$ with high accuracy is required for
the evaluation of 
the hadronic contribution $a_{\mu}^{had}$ to the anomalous magnetic
moment of the muon (g-2)/2 (see \cite{kino} and references 
therein). The ultimate goal of the
experiment E821 \cite{E821} running in Brookhaven National Laboratory
is to measure the muon anomalous magnetic
moment with a relative precision of 0.35 ppm. Within the Standard Model
(SM) the uncertainty of the
theoretical value of the leading order $a_\mu$   is dominated by the 
uncertainty of the hadronic contribution
$a_{\mu}^{had}$ calculated via the dispersion integral 
\begin{equation}
\label{dint}
a_\mu^{had} = 
\left( \dfrac{\alpha m_\mu}{3\pi} \right)^2
\int_{4m_\pi^2}^{\infty} \dfrac{R(s)K(s)}{s^2}ds=
 \dfrac{m_\mu^2}{12\pi^3} 
\int_{4m_\pi^2}^{\infty} \dfrac{\sigma(s)K(s)}{s}ds,
\end{equation}
where $K(s)$ is the QED kernel and $\sigma(s)$ is the cross
section of $e^+e^- \to$ hadrons.
The precision of the $a_{\mu}^{had}$ calculation depends on the
approach used and varies from 1.34 ppm based on $e^+e^-$ data only
\cite{EJ} to 0.53 ppm if in addition $\tau$-lepton decay data as well as 
perturbative QCD and QCD sum rules are extensively used \cite{DH}. 
The major contribution to its uncertainty comes from the 
systematic error of the $R(s)$ measurement at low energies
($s<$ 2 GeV$^2$), which, in turn, is dominated by the 
systematic error of the measured cross section \eepp .
 
Assuming conservation of the vector current (CVC) and 
isospin symmetry, the spectral function of the  
$\tau^-\to\pi^-\pi^0\nu_\tau$ decay can be related to the isovector
part of the pion form factor \cite{tsai}. The detailed measurement of 
the spectral functions was provided by ALEPH \cite{ALEPH}, 
OPAL \cite{OPAL} and CLEO-II \cite{CLEO}. The comparison of the pion 
form factor measured at
$e^+e^-$ colliders with the spectral function of the 
$\tau^-\to\pi^-\pi^0\nu_\tau$ decay provides a test
of CVC. If CVC holds with high accuracy,
$\tau$-lepton decay data can be also used to improve the accuracy
of the calculations mentioned above \cite{ADH,e20}.
 
E821 has recently published the result of its measurement of
$a_{\mu}$ with an accuracy of 1.3 ppm  \cite{BNL}.
The measured value of $a_{\mu}$ is 2.6 standard deviations 
higher than the SM prediction of \cite{DH}. \footnote{Recent progress
in estimating the light-by-light scattering contribution to
$a_{\mu}^{had}$ \cite{lbl} implies that the difference between
experiment and theory reduces to about 1.5 standard deviations.}  
This observation makes new high precision measurements of the
$e^+e^- \to hadrons$ cross section and particularly of the  
pion form factor extremely important.
  
Since early 70s the VEPP-2M $e^+e^-$ collider has been running in 
the Budker Institute of Nuclear Physics in the c.m. energy range 
360-1400 MeV. The most precise pion form factor data were
obtained in late 70s -- early 80s by CMD and OLYA detectors
\cite{OLYACMD}. Their accuracy was limited by systematic
errors of the experiments, varying from 2\% to 15\% over the VEPP-2M
energy range.

The CMD-2 detector 
installed in 1991 is a
general purpose detector consisting 
of the drift chamber, the proportional Z-chamber, the barrel CsI
calorimeter, the endcap BGO calorimeter
installed in 1996, and the muon range system. The 
drift chamber, Z-chamber and the endcap calorimeters are placed
inside a thin superconducting solenoid with a field of 1 T. 
More detail on the detector can be found elsewhere
\cite{PREP}. 

Though the collider luminosity has not considerably increased 
since the previous work in the 1980's, the present detector design 
is significantly improved, resulting in a smaller systematic error of 
the pion form factor in this measurement. 
Particularly, the following advantages of CMD-2 should be
mentioned compared to previous detectors: simultaneous
measurement of the particle momentum and energy deposition 
which simplified particle identification and helped to reduce background; 
high precision determination of the fiducial volume with the help
of the Z-chamber; a thinner beam pipe which reduced the nuclear 
interactions of pions.

During 1994-1995 a detailed scan of the c.m. energy range 
610-960 MeV was performed. Since the pion
form factor changes relatively fast in this energy range
dominated by the $\rho$-meson,
the systematic error due to an uncertainty in the energy
measurement can be significant. To reduce this contribution to
a negligible level, the beam energy was measured with the help of  
the resonance depolarization technique with an accuracy of 140 keV for
almost all energy points \cite{depol}. This paper presents the final 
analysis of the data taken in 1994-1995 with the integrated luminosity 
of about 310 nb$^{-1}$ and a systematic uncertainty of the 
cross section of 0.6\%. Preliminary results with a systematic uncertainty 
of 1.4\% were published in \cite{INP9910}, based on the same data sample.

\section{Data Analysis}

The data were collected at 43 points with c.m. energy ranging from 
610 MeV to 960 MeV in 10 MeV energy steps,
except for the narrow energy range near the
$\omega$-meson, where the energy steps were $2\div 6$ MeV.

From more than \( 4\cdot 10^{7} \) triggers recorded, 
about \( 3\cdot 10^{5} \) events 
were selected as \textit{collinear}, with a signature of two 
particles of opposite charge and nearly back-to-back momenta originating
from the interaction point. The following selection criteria were used: 
\label{collsel}
\begin{enumerate}
\item Two tracks of opposite charge originating from the interaction
region are reconstructed in the drift chamber. 
\item The distance from the vertex to the beam axis, $\rho$, is less than
0.3 cm and the z-coordinate of the vertex (along the beam axis) is
within $-15<z<15$ cm.
\item The average momentum of the two particles $(p_1+p_2)/2$ is 
between 200 and 600 MeV/c.
\item The difference between the azimuthal angles (in the plane
perpendicular to the beam axis) of two particles
$ |\Delta \varphi |=|\pi -|\varphi _{1}-\varphi _{2}||<0.15$. 
\item The difference between the polar angles (the angle between 
the momentum and the beam axis) of two particles
$|\Delta \Theta |=|\Theta _{1}-(\pi -\Theta _{2})|<0.25$.
\item The average polar angle of two particles $ \Theta_{avr}=[\Theta
_{1}+(\pi -\Theta _{2})]/2 $ is within $1.1<\Theta_{avr}<(\pi-1.1)$.
This criterion determines the fiducial volume.
\end{enumerate}

\begin{figure}
\begin{center}
\begin{tabular}{cc}
\includegraphics[width=0.45\textwidth]{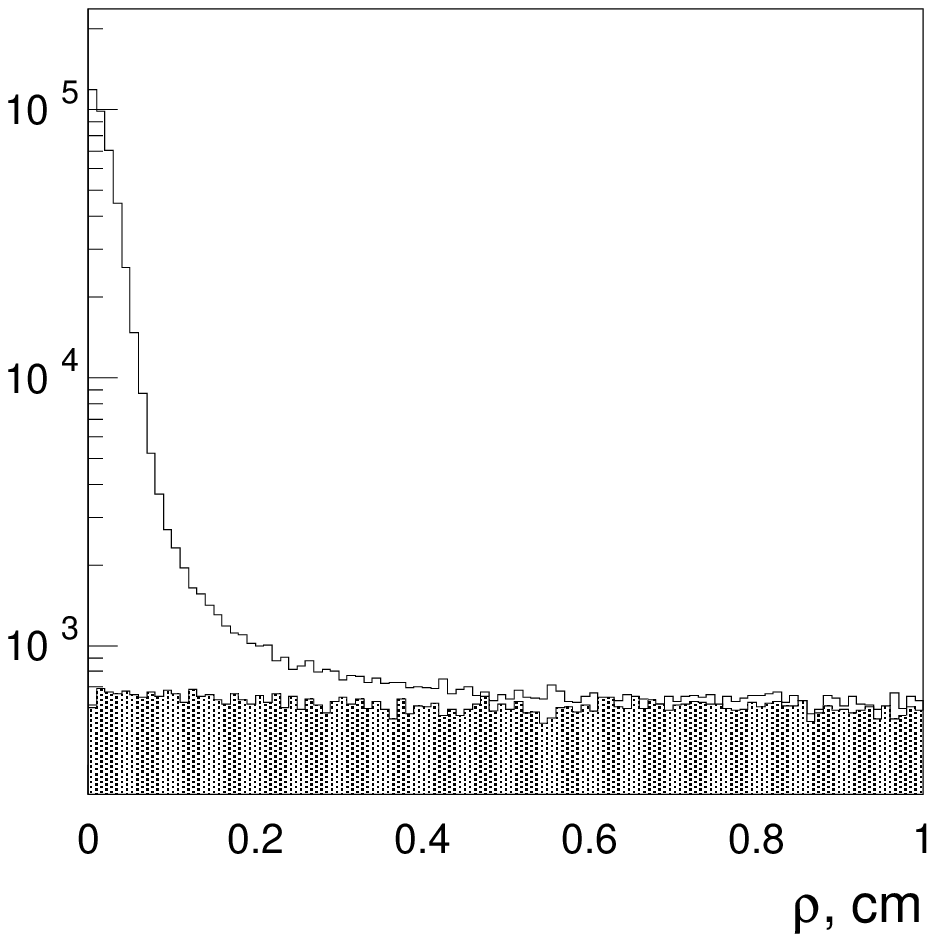} &
\includegraphics[width=0.45\textwidth]{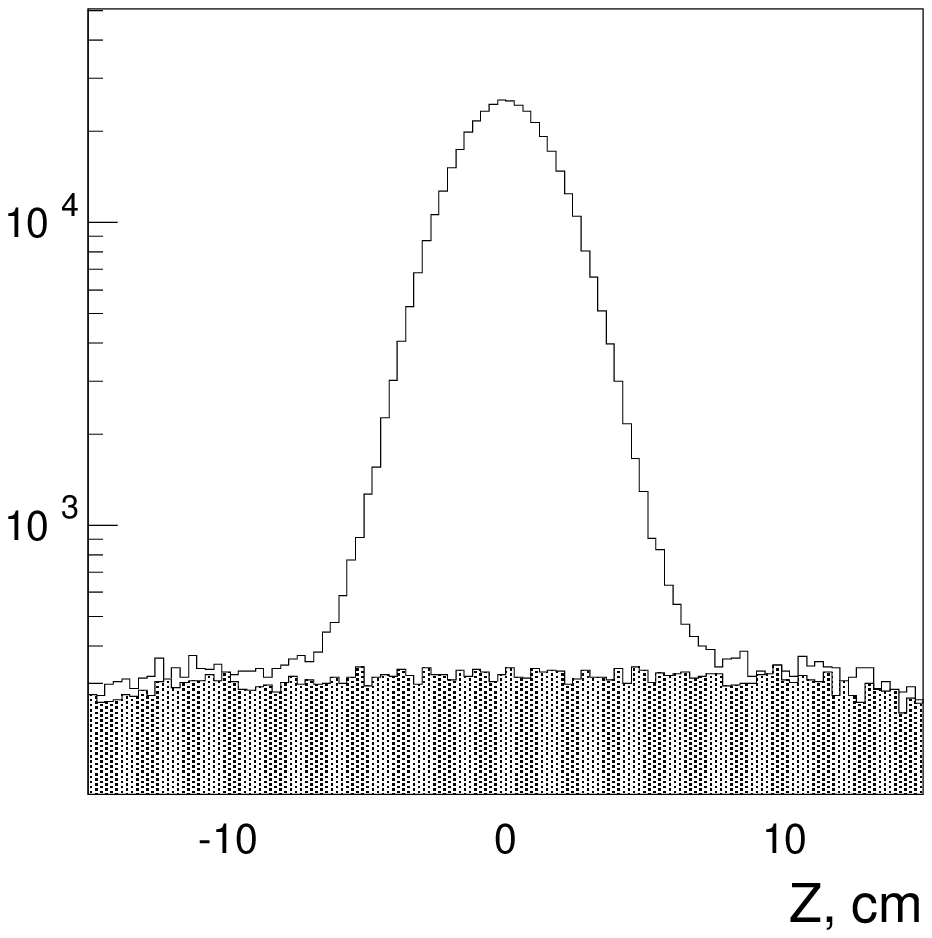} \\
\end{tabular}
\end{center}
\caption{\label{rhoz12}
Spatial distribution of the vertex. The left plot shows the 
distribution of the distance from the vertex to the beam axis
($\rho$), the right plot presents the distribution of the distance 
between the vertex and the center of the
detector along the beam axis ($z$). 
The open histogram corresponds to all collinear events, the filled one
shows the subset of background events.}
\end{figure}

The selected sample of collinear events contains 
$e^+e^-\to e^+e^-$, 
$e^+e^-\to\pi^+\pi^-$, $e^+e^-\to\mu^+\mu^-$ events (below referred to as
beam originating) as well as the background of cosmic particles
which pass near the interaction region and  are misidentified as 
collinear events.
The number of cosmic background events $N_{cosmic}$ was 
determined by the analysis of the spatial distribution of the vertex. 
Both distributions of the longitudinal coordinate ($z$) and the distance
from the beam axis ($\rho$) are peaked around zero for the
beam originating events, but are very broad, 
almost flat for the cosmic background events. Typical
$\rho$- and $z$-distributions are shown in Fig. \ref{rhoz12}.

The momenta of $e$, $\mu$, $\pi$ from the beam originating events are
rather close (the difference is comparable to the momentum
resolution of the drift chamber), so the overlap 
of the momentum distributions is large. Therefore, separation of
$e$, $\mu$ and $\pi$ by their momenta is impossible at 
$\sqrt{s} >$ 600 MeV. 

\begin{figure}
\begin{center}
\includegraphics[width=0.45\textwidth]{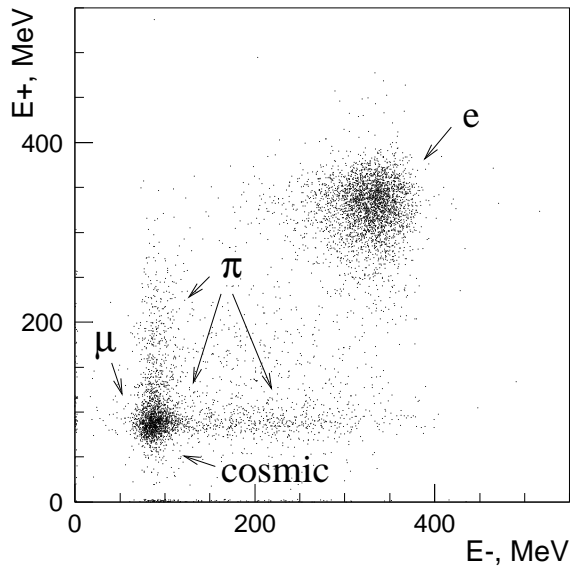}
\end{center}
\caption{\label{distr}Energy deposition of collinear events for the 
beam energy  of 400 MeV.}
\end{figure}

On the contrary, the energy deposition of the particles in the 
calorimeter is quite different for $e$, $\mu$ and $\pi$. The typical 
energy deposition of two particles ($E^+$ vs $E^-$) for
experimental  events selected at the beam energy of 400 MeV  
is shown in Fig.\ \ref{distr}. The high 
deposition spot corresponds 
to $e^+e^-$ pairs, where both particles leave almost all their 
energy in the calorimeter. The low deposition spot represents 
$\mu^+\mu^-$ pairs, cosmic muons and those $\pi^+\pi^-$ pairs, 
in which both particles interact as minimum-ionizing. The long tails 
correspond to $\pi^+\pi^-$ pairs in which one or both particles 
undergo nuclear interactions inside the calorimeter. 

Therefore, the energy deposition of the particles was used for the 
separation of the beam originating events. Since the overlap of the 
distributions for $e^+e^-$ and $\pi^+\pi^-$ pairs is small, this approach 
gives stable results with a small systematic error. However,
$\mu^+\mu^-$ and $\pi^+\pi^-$ pairs cannot be separated well by 
their energy deposition. To avoid this problem, the number of 
$\mu^+\mu^-$ pairs was derived
from the number of $e^+e^-$ pairs according to QED,
taking into account radiative corrections and detection efficiencies.
Since in this energy range the number of $\mu^+\mu^-$ pairs is 
small compared to that  of $\pi^+\pi^-$, the systematic error 
caused by the corresponding calculation 
is negligible (less than 0.03\%).

The separation was based on the minimization of the following unbinned
likelihood function:
\begin{equation}
\label{lfunc1}
L= - \sum_{events} \ln \left( \sum_a N_a\cdot f_a(E^+,E^-) \right) +
\sum_a N_a,
\end{equation}
where $a$ is the event type ($a=ee$, $\mu\mu$, $\pi\pi$, $cosmic$),
$N_a$ is the number of events of the type $a$ and $f_a(E^+,E^-)$
is the probability density function (p.d.f.) for a type $a$ event to
have energy depositions $E^+$ and $E^-$. It was assumed that $E^+$ 
and $E^-$ are not correlated for events of the same type,
so the p.d.f. can be factorized as
\[ f_a(E^+,E^-)=f_a^+(E^+)\cdot f^-_a(E^-), \]
where $f^{\pm}_a(E)$ is the p.d.f.\ for 
$e^\pm$, $\mu^\pm$, $\pi^\pm$ and cosmic muons to have 
energy deposition equal to $E$.
This assumption is not entirely correct since
there are small correlations between $E^+$ and $E^-$ because of
the dependence of the calorimeter thickness on the polar angle 
as well as the c.m. energy shift due to the initial state radiation. 
The first effect was corrected for while the second one was studied 
with the help of simulation and was shown to be negligible (below 0.1\%). 

For $e^+e^-$, $\mu^+\mu^-$ pairs and cosmic events the
energy deposition does not depend on the particle charge,
while the energy depositions for $\pi^+$ and $\pi^-$ are different.
Therefore $f_a^+\equiv f^-_a$ for $a=ee,\; \mu \mu$
and $cosmic$, but in case of pions the corresponding p.d.f.'s\ 
are different. 

As was mentioned before, the ratio $N_{\mu\mu}/N_{ee}$ was fixed
during minimization according to the QED calculation
\[
\frac{N_{\mu\mu}}{N_{ee}}=
\frac{\sigma_{\mu\mu}\cdot (1+\delta_{\mu\mu})
 \varepsilon_{\mu\mu}}
{\sigma_{ee}\cdot (1+\delta_{ee}) 
\varepsilon_{ee}},
\]
 where $\sigma$ is the Born cross section, $\delta$ is the
radiative correction and \( \varepsilon  \) is the detection
efficiency which includes acceptance as well as  reconstruction
and trigger efficiencies. 
The number of cosmic events $N_{cosmic}$ determined separately
was fixed during the minimization while its fluctuation was 
added in quadrature to the fluctuation of $N_{\pi\pi}$. 

To obtain the specific form of p.d.f.'s, the energy
deposition of $e$, $\mu$ and $\pi$ in the CMD-2 calorimeter was
studied. 
For electrons (positrons) and cosmic muons it can be obtained 
with the help of the data itself. Particles of positive
charge with large enough energy deposition are almost 100\%
positrons and they were used to tag electrons. Events with a large
value of $\rho$ are mostly
cosmic muons. Such tagged particles were used to determine  the energy
deposition of electrons and cosmic muons respectively. 

The simulation was used to obtain the energy deposition of muons 
from $e^+e^-\to\mu^+\mu^-$. In the energy range under study
these muons interact purely as minimum-ionizing particles which
are well described by the simulation. 

On the contrary, the simulation of the interaction of low energy pions
inside the calorimeter is not reliable enough. In addition, there is no 
good method to tag pions from $e^+e^-\to\pi^+\pi^-$ events. Therefore,
the p.d.f.'s for $\pi^+$ and $\pi^-$ were obtained from the analysis 
of the energy deposition of pions coming from the 
$\phi(1020)\to\pi^+\pi^-\pi^0$ decay. From a large data 
sample collected by CMD-2 around the $\phi$-meson peak, about 10$^5$ 
$\phi(1020)\to 3\pi$ events were selected with practically no background. 
Pions found in these events cover the whole interesting range of 
momenta and angles. The energy deposition of these pions was analyzed 
and the parameterization for $\pi^+$ and $\pi^-$ p.d.f.'s was derived.

Finally, to simplify the final error calculation, the 
likelihood function (\ref{lfunc1}) 
was rewritten to have the following global fit parameters:
\[ (N_{ee}+N_{\mu\mu}),\quad \frac{N_{\pi\pi }}{N_{ee}+N_{\mu\mu}} \]
instead of $N_{ee}$ and  $N_{\pi\pi}$ (with $N_{\mu\mu}/N_{ee}$ and
$N_{cosmic}$ fixed). The likelihood function has some other 
fit parameters characterizing the
p.d.f.'s\ for different types of particles,
such as the mean energy, energy resolution, the asymmetry of
some distributions etc. 
More detail about the energy deposition of different types of
collinear events and the p.d.f.\ parameterization can be found in
\cite{INP9910}. 

After the separation procedure the following number    
of events was obtained for three described above classes:
 $N_{ee}+N_{\mu\mu}=180038, N_{\pi\pi}=113824$ and $N_{cosmic}=17390$.

\begin{figure}
\begin{center}
\includegraphics[width=0.45\textwidth]{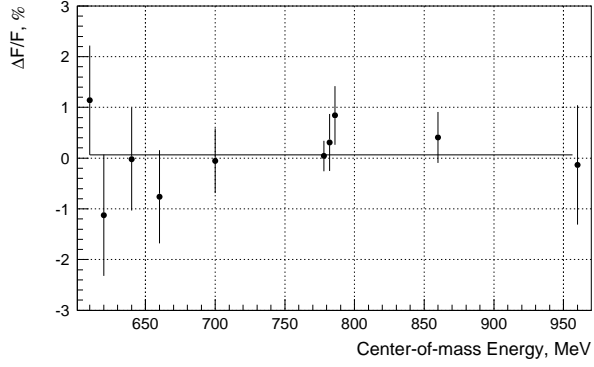}
\end{center}
\caption{\label{fig:sepsim}Difference between the initial and the
reconstructed values of the  form factor for the simulated data set.} 
\end{figure}

Special studies were performed to estimate the systematic error of
the separation procedure. The dominant effect was produced by the 
small nonuniformity of the calorimeter calibration.
Due to the forward-backward asymmetry of the  \eeee  cross section,
the calorimeter calibration error leads to a small difference between 
$e^+$ and $e^-$ energy 
depositions. The corresponding error was found to be less than 0.2\%.
Several different functional forms were used to parameterize p.d.f.'s 
of $e$'s and $\pi$'s and the final cross section was stable within 0.1\% 
for different selections.
The existing variation of the calorimeter response between
calibrations leads to small variations of the energy resolution 
which could also influence the results. The estimated contribution 
of this effect is below 0.1\%.
As a final test, the large amount of $e^+e^-\to e^+e^-(\gamma)$, 
$\mu^+\mu^-(\gamma)$, $\pi^+\pi^-(\gamma)$ events was generated in a 
proper proportion with the help of full detector 
simulation at several energy points covering the whole energy range.
After that the simulated events were subject to the same
separation procedure as for the data. Results are shown in 
Fig.\ \ref{fig:sepsim}. The
reconstructed value is always consistent with the input one and the
average difference is below 0.1\% or consistent with
zero. Thus, the systematic error because of the event separation 
taking into account the above effects
is estimated to be 0.2\%.

\section{Form factor calculation}

The pion form factor was calculated as:
\begin{equation}
\label{piform}
|F_\pi|^2 = \frac{N_{\pi\pi}}{N_{ee}+N_{\mu\mu}} \times 
\frac{
\sigma_{ee}\cdot(1+\delta_{ee})\varepsilon_{ee} +
\sigma_{\mu\mu}\cdot(1+\delta_{\mu\mu})\varepsilon_{\mu\mu}
}{\sigma_{\pi\pi}\cdot(1+\delta_{\pi\pi})
(1+\Delta_N)(1+\Delta_D)\varepsilon_{\pi\pi}}
-\Delta_{3\pi},
\end{equation}
where the ratio $N_{\pi\pi}/(N_{ee}+N_{\mu\mu})$ was obtained from the
minimization of  (\ref{lfunc1}),
$\sigma$ are the corresponding Born cross-sections,
$\delta$ are the radiative corrections, $\epsilon$ are the detection 
efficiencies, 
$\Delta_D$ and $\Delta_N$ are the corrections for the pion losses
caused by decays in flight and nuclear interactions respectively,
and $\Delta_{3\pi}$  is the correction for misidentification of
$\omega\to\pi^+\pi^-\pi^0$ events as the \eepp. In the case of the
\eepp process, $\sigma_{\pi\pi}$ corresponds to pointlike pions.

Radiative corrections  were calculated according to Refs. 
\cite{rcee,rcpp} in which the accuracy of the obtained formulae was
0.2\%. \footnote{We corrected a 
misprint in formula (2.5) of Ref.\  \cite{rcpp}, also noted 
by the authors 
of Ref.\ \cite{hoefer}.} 
The improved precision of the calculations compared to previous
works \cite{ber} comes from taking into account 
the radiation of the additional photon in a narrow
cone  along the direction of electrons and positrons.
The radiative corrections   were calculated by the Monte Carlo
integration of the differential cross sections imposing 
all selection criteria.
 We estimate the uncertainty  of the
form factor because of the radiative corrections to be 0.4\%
dominated by the accuracy of the ratio
$(1+\delta_{ee})/(1+\delta_{\pi\pi})$. 

The radiative corrections for the process \eepp 
include the effects of both initial (ISR) and final state radiation 
(FSR) and do not include the vacuum polarization terms
(both leptonic and hadronic) since the latter are considered
to be an  intrinsic part of the hadronic cross section and
corresponding form factor \eqref{ffdef}.
However, for
various applications based on dispersion relations and involving
the total cross section of $e^+e^- \to$ hadrons, the radiation
by final pions is no longer a radiative correction, so that 
$\pi^+\pi^-\gamma$ with a photon radiated by one of the final pions 
should be considered as one of the possible hadronic final states 
contributing to the total cross section. Therefore, the 
bare cross section $\eepp(\gamma)$, below referred to as 
$\sigma^0_{\pi\pi(\gamma)}$, was also calculated as 
\begin{equation}
\label{bare}
\sigma^0_{\pi\pi(\gamma)} = 
\frac{\pi\alpha^2}{3s}\beta^3_\pi \left| F_\pi(s)\right|^2
\cdot |1-\Pi(s)|^2 
\cdot \left( 1+\frac{\alpha}{\pi}\Lambda(s) \right).
\end{equation}
The factor $|1-\Pi(s)|^2$ with a polarization operator 
$\Pi(s)$ excludes the effect of leptonic and hadronic vacuum 
polarization, so that one obtains the bare cross section required
for various applications. \footnote{ Note that the  
definition of the cross section above is different from 
that in \cite{INP9910} where the FSR cross section was not taken into
account and only the correction for the leptonic vacuum polarization
was applied. Therefore, their comparison is meaningless.} 
A correction for the final state radiation 
$\Lambda(s)$ was calculated based on \cite{rcpp}, where only the
effects of FSR, integrated over the whole allowed kinematic region,
were taken into account. The result 
\begin{equation}
\begin{array}{ll}
\Lambda(s)  = &  \frac {1+\beta_{\pi}^2} {\beta_{\pi}} 
\{ 4\text{Li}_2 (\frac {1-\beta_{\pi}} {1+\beta_{\pi}})+
2\text{Li}_2 (-\frac {1-\beta_{\pi}} {1+\beta_{\pi}})
 - [3\ln(\frac{2}{1+\beta_{\pi}})
+ 2\ln\beta_{\pi} ]L_{\beta{\pi}}\} \\
 & -  3 \ln  (\frac {4}{1-\beta_{\pi}^2})-4\ln\beta_{\pi} 
   +  \frac {1} {\beta_{\pi}^3}
 \left[\frac{5}{4}(1+\beta_{\pi}^2)^2-2\right]L_{\beta_{\pi}} 
 + \frac {3} {2} \frac{1+\beta_{\pi}^2}{\beta_{\pi}^2},
\end{array}
\end{equation}
where   $L_{\beta_{\pi}} = 
\ln \frac{1+\beta_{\pi}}{1-\beta_{\pi}}, 
 \quad \text{Li}_2(z) = -\int_0^z \frac{dx}{x}\ln(1-x)$,
coincides with \cite{kirill,hoefer}.

Two different trigger settings were used during data
taking. For energies below 810 MeV only a charged 
trigger was used in which the positive decision was based 
on the information from the  tracking system only.  
The efficiency of the charged
trigger was measured to be  higher than 99\% and, more important, 
equal for different types of collinear events in the energy range under 
consideration. 
Therefore, the uncertainties related to the trigger efficiency
cancel in \eqref{piform}.
Above 810 MeV the additional
requirement that the energy deposition  in the
calorimeter be at least 20 MeV was applied for the trigger. The efficiency of
this requirement was measured to be 99.5\% for \eepp, 99.2\% for \eemm
and 100\% for \eeee events, and was included in the corresponding
detection efficiency. These corrections give a negligible contribution
to the systematic uncertainty.

The reconstruction efficiency was measured using the experimental data
themselves. It was found to be within the 98-100\% range at all energies
and, within the  statistical accuracy, nearly the same for all types of
collinear events. Therefore, it cancels 
in \eqref{piform}. 
The systematic error of the cancelation was
estimated to be better than 0.2\%. It's worth noting that such
a cancelation allows a determination of the form factor with better
precision than that of the luminosity.  

The fiducial volume (detection solid angle) is determined by
selecting events with the average polar angle
$\Theta_{avr}=(\Theta_1+\pi-\Theta_2)/2$ in the range between
$\Theta_{min}$ and $(\pi-\Theta_{min})$ with $\Theta_{min}=1.1$. 
The value of $\Theta_{avr}$ is determined by the CMD-2
Z-chamber \cite{ZC} which has a spatial resolution along the beam
axis better than 0.6 mm. That corresponds to 
a systematic error in the form factor of about
0.2\%. To test this estimate, the pion form
factor was also determined at $\Theta_{min}$ = 1.0 radian. 
The difference between the form factor values determined at
two values of $\Theta_{min}$ averaged over the c.m. energy range
was found to be $(0.1 \pm 0.3)$\%, consistent with zero.
 
The correction $\Delta_D$ for the pion losses caused by 
decays in flight was calculated with the help of simulation. Its
value, varying from 0.2\% at 600 MeV
to 0.03\% around the $\phi(1020)$-meson, turned out to be small
because of the 
small size of the decay volume and the fact that the maximum decay 
angle is small enough and of the same magnitude as the angular 
resolution of the drift chamber. This effect gives a negligible 
contribution to the systematic uncertainty. 

The correction $\Delta_N$ for the pion losses caused by nuclear
interactions inside the wall of the beam pipe or the drift chamber
was calculated with the help of FLUKA-based simulation \cite{FLUKA}.
The value of the correction is slowly changing from 1.7\% at 600 MeV
to 0.8\% at the $\phi(1020)$-meson energy. The systematic error of 
0.2\% for the $\Delta_N$ calculation was estimated from the 
uncertainty of the nuclear cross sections used in FLUKA \cite{nucl}.

There is also a small correction for the losses of
\eeee events, caused by interactions of electrons with
material of the beam pipe and the drift chamber. It was taken into
account by reducing $\Delta_N$ by 0.15\% according to the
simulation.

Use of the resonance depolarization for the absolute beam energy
calibration reduced the systematic uncertainty of the form factor
due to the beam energy measurement to 0.1\%.
 
In the narrow energy range around the $\omega(782)$-meson there is a small
background of $\ee\to\pi^+\pi^-\pi^0$ events, misidentified as \eepp .
The corresponding correction 
$\Delta_{3\pi}$ was calculated from the simulation. It reaches
its maximum value of about 1\% at the $\omega(782)$-meson energy and drops 
fast to nearly zero at the energies outside  the $\omega$-meson.

The measured values of the pion form factor as well as those of the
bare $\eepp(\gamma)$ cross-section obtained from \eqref{bare} are shown 
in Table \ref{table:fpi}.  
Only statistical errors are given.
The main sources of the systematic error are listed in 
Table \ref{tab:syst}.
The overall systematic error obtained by summing individual 
contributions in quadrature is about 0.6\%.
The values of the pion form factor and the bare cross section in
Table \ref{table:fpi}  supersede our preliminary results 
presented in Table 3 of Ref.~\cite{INP9910}. 

\begin{table*}
\caption{\label{table:fpi} The measured values of the pion form
factor and bare cross section $\eepp(\gamma)$. 
Only statistical errors are shown. The systematic error is
estimated to be 0.6\%.}
\begin{tabular}[t]{ccc@{\hspace{10mm}}ccc}
\hline 
2E (MeV) & $|F_\pi|^2$ & $\sigma^0_{\pi\pi(\gamma)}$, nb &
2E (MeV) & $|F_\pi|^2$ & $\sigma^0_{\pi\pi(\gamma)}$, nb \\
\hline 
$610.50$ & $ 8.00 \pm 1.13$ & $ 327.8 \pm 46.2$ & $784.24$ & $34.30 \pm 0.97$ & $ 935.9 \pm 26.5$ \\ 
$620.50$ & $ 9.71 \pm 0.72$ & $ 390.6 \pm 29.1$ & $786.04$ & $30.01 \pm 1.07$ & $ 809.5 \pm 28.9$ \\ 
$630.50$ & $10.82 \pm 0.73$ & $ 426.9 \pm 28.6$ & $790.10$ & $31.96 \pm 1.11$ & $ 859.3 \pm 30.0$ \\ 
$640.51$ & $11.02 \pm 0.72$ & $ 426.2 \pm 27.9$ & $794.14$ & $29.81 \pm 0.85$ & $ 799.6 \pm 22.7$ \\ 
$650.49$ & $12.66 \pm 0.83$ & $ 480.3 \pm 31.6$ & $800.02$ & $29.40 \pm 0.68$ & $ 782.7 \pm 18.1$ \\ 
$660.50$ & $13.40 \pm 0.75$ & $ 498.8 \pm 27.8$ & $810.14$ & $25.60 \pm 0.55$ & $ 669.7 \pm 14.4$ \\ 
$670.50$ & $15.36 \pm 0.82$ & $ 560.5 \pm 30.0$ & $820.02$ & $24.09 \pm 0.78$ & $ 619.0 \pm 20.0$ \\ 
$680.59$ & $19.49 \pm 0.90$ & $ 697.2 \pm 32.3$ & $829.97$ & $20.52 \pm 0.73$ & $ 517.5 \pm 18.5$ \\ 
$690.43$ & $20.49 \pm 0.73$ & $ 718.6 \pm 25.5$ & $839.10$ & $16.75 \pm 0.70$ & $ 415.4 \pm 17.3$ \\ 
$700.52$ & $23.81 \pm 0.62$ & $ 818.1 \pm 21.2$ & $849.24$ & $14.16 \pm 0.70$ & $ 344.8 \pm 17.0$ \\ 
$710.47$ & $27.52 \pm 0.99$ & $ 926.1 \pm 33.4$ & $859.60$ & $14.47 \pm 0.67$ & $ 345.7 \pm 15.9$ \\ 
$720.25$ & $31.80 \pm 0.82$ & $1048.1 \pm 27.0$ & $869.50$ & $11.10 \pm 0.46$ & $ 260.7 \pm 10.7$ \\ 
$730.24$ & $34.34 \pm 1.18$ & $1107.2 \pm 38.1$ & $879.84$ & $10.13 \pm 0.79$ & $ 233.6 \pm 18.1$ \\ 
$740.20$ & $37.82 \pm 1.18$ & $1191.9 \pm 37.1$ & $889.72$ & $ 8.44 \pm 0.34$ & $ 191.1 \pm  7.7$ \\ 
$750.28$ & $42.33 \pm 1.18$ & $1302.9 \pm 36.2$ & $900.04$ & $ 7.74 \pm 0.29$ & $ 172.1 \pm  6.4$ \\ 
$760.18$ & $43.44 \pm 1.21$ & $1306.9 \pm 36.5$ & $910.02$ & $ 6.82 \pm 0.31$ & $ 149.1 \pm  6.8$ \\ 
$764.17$ & $43.23 \pm 1.03$ & $1289.6 \pm 30.6$ & $919.56$ & $ 6.03 \pm 0.30$ & $ 129.7 \pm  6.5$ \\ 
$770.11$ & $44.23 \pm 1.14$ & $1304.6 \pm 33.5$ & $930.11$ & $ 5.70 \pm 0.36$ & $ 120.3 \pm  7.7$ \\ 
$774.38$ & $43.19 \pm 1.15$ & $1265.7 \pm 33.7$ & $942.19$ & $ 5.16 \pm 0.25$ & $ 106.8 \pm  5.1$ \\ 
$778.17$ & $45.41 \pm 1.29$ & $1321.6 \pm 37.7$ & $951.84$ & $ 4.56 \pm 0.23$ & $  92.9 \pm  4.8$ \\ 
$780.17$ & $42.90 \pm 1.21$ & $1234.9 \pm 34.8$ & $961.52$ & $ 4.29 \pm 0.23$ & $  86.2 \pm  4.6$ \\ 
$782.23$ & $36.59 \pm 0.66$ & $1026.6 \pm 18.5$ & \\ 

\hline
\end{tabular}
\end{table*}

\begin{table}
\caption{\label{tab:syst} Main sources of the systematic errors}
\begin{tabular}{lc}
\hline
Source & Contribution \\
\hline 
Event separation& 0.2\% \\
Radiative corrections & 0.4\% \\
Detection efficiency & 0.2\% \\
Fiducial volume &  0.2\% \\
Correction for pion losses & 0.2\% \\
Beam energy determination & 0.1\% \\
\hline
Total & 0.6\% \\
\hline 
\end{tabular}
\end{table}

\section{Fit to data}

The parameterization of the pion form factor in the energy range under
study should include contributions from the $\rho(770)$, $\omega(782)$
and $\rho(1450)$ resonances. 
It was  assumed that
the only mechanism for the $\omega\to\pi^+\pi^-$ decay is $\rho-\omega$
mixing. Following \cite{Feynman}, we represent the wave function of
the $\omega(782)$-meson as 
\[\label{omega}
|\omega\rangle=|\omega_0\rangle+\varepsilon|\rho_0
\rangle\, ,
\]
where $|\omega_0\rangle$ and $|\rho_0\rangle$ are the pure isoscalar and
isovector respectively, and $\varepsilon$ is the $\rho-\omega$ mixing
parameter. Then, in the energy region close to the $\rho(770)$- and 
$\omega(782)$-meson masses, the form factor can be written as
\begin{equation}\label{FF}
F_{\pi}(s)=\left[ \frac{F_{\rho}}{s-M_{\rho}^2}
+\varepsilon\,\frac{F_{\omega}}{s-M_{\omega}^2} \right]
\left[ \frac{F_{\rho}}{-M_{\rho}^2}
+\varepsilon\,\frac{F_{\omega}}{-M_{\omega}^2} \right]^{-1}
\approx -\frac{M_{\rho}^2}{s-M_{\rho}^2}
\left[1+\varepsilon \, 
\frac{F_{\omega}(M^2_{\omega}-M^2_{\rho})s}
{F_{\rho}M_{\omega}^2(s-M_{\omega}^2)}\,\right]\, ,
\end{equation}
where we keep only the terms linear in $\varepsilon$. The quantities
$M_{\omega}$ and $M_{\rho}$ are complex and contain the
corresponding widths.   

Combining together the contributions from the $\rho(770)$- and 
$\rho(1450)$-mesons, and including that from the $\omega(782)$-meson as in 
\eqref{FF}, we write the pion form factor:
\begin{equation}
\label{func:GS}
F_\pi(s)=\frac{\GS_{\rho(770)}(s)\cdot
\left( {\mathstrut 1+\delta \,
\frac{\displaystyle s}{\displaystyle M_\omega^2}} \, \BW_{\omega}(s) \right)
+ \beta \cdot \GS_{\rho(1450)}(s)
}{1+\beta},
\end{equation}
where parameters $\delta$ and $\beta$ describe the
contributions of the $\omega(782)$- and $\rho(1450)$-mesons relative to the
dominant one of the $\rho(770)$-meson. 
For the $\rho(770)$ and $\rho(1450)$ 
the GS parameterization is used~\cite{GS}:
\begin{equation}
\GS_{\rho(\mrho)} = \frac{
\mrho^2 \left( 1 + d \cdot \Gamma_\rho / \mrho \right)
}{
\mrho^2 - s + f(s) - i \mrho \Gamma_\rho(s)
}, \quad \text{where}
\end{equation}
\begin{equation}
f(s) = \Gamma_\rho \frac{\mrho^2}{p_\pi^3(\mrho^2)}
\left[
p_\pi^2(s) \left( h(s) - h(\mrho^2) \right) + 
(\mrho^2-s) \, p_\pi^2(\mrho^2) 
\left. \frac{dh}{ds} \right|_{s=\mrho^2}
\right] ,
\end{equation}
\begin{equation}
h(s) = \frac{2}{\pi} \frac{p_\pi(s)}{\sqrt{s}}
\ln \frac{\sqrt{s}+2 p_\pi(s)}{2m_\pi} ,
\end{equation}
and $d$ is chosen to satisfy $\GS_{\rho(\mrho)}(0)=1$:
\begin{equation}
d = \frac{3}{\pi} \frac{m_\pi^2}{p_\pi^2(\mrho^2)}
\ln \frac{\mrho + 2 p_\pi(\mrho^2)}{2m_\pi}
+ \frac{\mrho}{2\pi p_\pi(\mrho^2)}
- \frac{m_\pi^2 \mrho}{\pi p_\pi^3(\mrho^2)},
\end{equation}
where $\mrho, \Gamma_{\rho}$ are the $\rho(770)$-meson mass and width,
and $p_{\pi}(s)$ is the pion momentum at the squared 
c.m. energy s. For the energy dependence of the $\rho(770)$ width, the 
P-wave phase space is taken:
\begin{equation}
\label{rhowidth}
\Gamma_\rho(s) = \Gamma_\rho 
\left[ \frac{p_\pi(s)}{p_\pi(\mrho^2)} \right]^3
\left[ \frac{\mrho^2}{s} \right]^{1/2}.
\end{equation}
For the $\omega(782)$-meson contribution the simple Breit-Wigner
parameterization with a constant width is used. 

To obtain the $\rho(770)$-meson leptonic width $\Gamma(\rho\to
e^+e^-)$, the well known VDM relations were used \cite{Newrho}:
\begin{equation}
\label{Gvee}
\Gamma_{V\rightarrow e^+e^-} = 
\frac{4\pi\alpha^2}{3\mathrm{M}_V^3} g_{V\gamma}^2
\quad \text{and} \quad
\Gamma_{V\rightarrow\pi^+\pi^-} = 
\frac{g_{V\pi\pi}^2}{6\pi} 
\frac{p_\pi^3 \left( \mathrm{M}_V^2 \right)}{\mathrm{M}_V^2}.
\end{equation}
Assuming 
\begin{equation}
g_{\rho\gamma} g_{\rho\pi\pi} = 
\frac{ \mrho^2 \left( 1 + d \cdot \Gamma_\rho / \mrho \right) }
{(1+\beta)}
\quad \text{and} \quad \Gamma_{\rho\to\pi^+\pi^-}=\Gamma_\rho,
\end{equation}
the following result was obtained: 
\begin{equation}
\Gamma_{\rho\rightarrow e^+e^-} = 
\frac{2\alpha^2 p_\pi^3 \left( \mrho^2 \right)}{9 \mrho \Gamma_\rho}
\frac{(1+ d \cdot \Gamma_\rho / \mrho )^2}{(1+\beta)^2}.
\end{equation}

Similarly, assuming \eqref{Gvee} and 
\begin{equation}
g_{\omega\gamma} g_{\omega\pi\pi} = 
\frac{ \delta \cdot \momg^2 \cdot |\GS_{\rho(770)}(\momg^2)| }
{(1+\beta)},
\end{equation}
the following expression is obtained for the branching ratio of
the $\omega\to\pi^+\pi^-$ decay: 
\begin{equation}
Br(\omega\rightarrow\pi^+\pi^-) = 
\frac{2\alpha^2 p_\pi^3(\momg^2)}
{9\momg\Gamma_{\omega\rightarrow e^+e^-}\Gamma_\omega}
\left| \GS_{\rho(770)}(\momg^2) \right|^2
\frac{|\delta|^2}{(1+\beta)^2}.
\end{equation}

It should be mentioned that our 
parameterization of the $\omega(782)$-meson contribution 
is slightly different from that in \cite{ALEPH,OLYACMD}:
$(1+\delta\cdot s/M_\omega^2 \cdot \BW_{\omega}(s) )$ instead of
$(1+\delta \cdot \BW_{\omega}(s) ) / (1+\delta)$.
Fits to either parameterization give the same 
result.

To estimate the model dependence of the obtained parameters,
another model of the form factor parameterization was considered -
the Hidden Local Symmetry (HLS) model \cite{Newrho,HLS}. In this model 
the $\rho(770)$-meson appears as a dynamical gauge boson of a hidden local 
symmetry in the non-linear chiral Lagrangian. 
The $\rho(1450)$ contribution is not taken into
account, replaced by the non-resonant coupling $\gamma\pi^+\pi^-$. 
This model introduces a
real parameter $a$ related to this non-resonant coupling.
The original parameterization of the pion form factor was modified in
a similar way to
take into account the $\omega(782)$-meson contribution.

\section{Results and discussion}

\subsection{$\rho(770)$- and $\omega(782)$-meson parameters}

There are several parameters of both models whose values have to be
taken from other measurements \cite{PDG,omega}. Their values  
were allowed to fluctuate within the stated experimental errors. 
The following values of the $\omega(782)$-meson parameters were taken
from the CMD-2 experiment \cite{omega}: 
$\momg=(782.71\pm 0.08)$ MeV,
$\Gamma_\omega=(8.68\pm 0.24)$ MeV, $\Gamma_{\omega ee}=(0.595\pm 0.017)$
keV. Parameters of the $\rho(1450)$ were taken from \cite{PDG}:
$\mathrm{M}_{\rho(1450)}=(1465\pm 25)$ MeV and
$\Gamma_{\rho(1450)}=(310\pm 60)$ MeV.

\begin{figure}
\begin{center}
\includegraphics[width=0.45\textwidth]{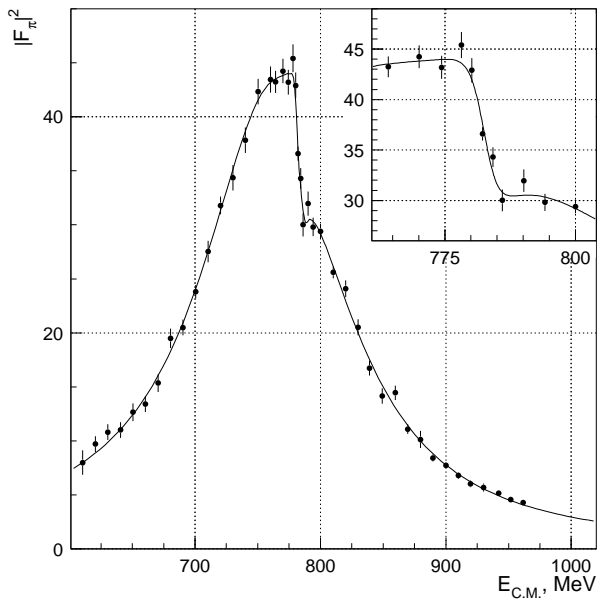}
\end{center}
\caption{\label{fpi}Fit of the CMD-2 pion form factor data to the
GS model.}
\end{figure}

Results of the fit in the GS model are shown in Fig.\ \ref{fpi}.  
Parameters of GS and HLS models obtained from the fit are in good
agreement with each other and 
are shown in Table \ref{rhopar}. The first error is 
statistical, the second one
is systematic. Two effects were taken into account in  the 
estimation of the latter: the systematic uncertainty of the form factor
measurement of 0.6\% (Table \ref{tab:syst}) and the contribution of 
the $\rho(1700)$ resonance missing in the adopted GS
parameterization. 
The parameter $\beta$ effectively describes 
the overall contribution of the $\rho(1450)$ and $\rho(1700)$
and its value is strongly correlated with the $\rho(1450)$ width.
To extract the $\rho(1450)$ contribution, a fit of the data should 
be performed in a broader energy range. As a result, the value
of $\beta$ is strongly model dependent and is not well defined 
in the energy range under study. For this reason, its systematic
error is not shown. 

\begin{table}
\caption{\label{rhopar}Results from fits to $|F_\pi(s)|^2$ for the GS
and HLS models}
\vspace{1mm}
\begin{tabular}{lcc}
\hline
Parameter & GS model & HLS model \\ \hline
$\mrho$, MeV & $776.09\pm 0.64\pm 0.50$ & $775.23\pm 0.61\pm 0.50$ \\
$\Gamma_\rho$, MeV & $144.46\pm 1.33 \pm 0.80$ & $143.88\pm 1.44\pm
0.80$  \\
$\Gamma(\rho\rightarrow e^+e^-)$, keV & $6.86\pm 0.11\pm 0.05$ &
$6.84\pm 0.12\pm 0.05$ \\
$Br(\omega\rightarrow\pi^+\pi^-)$, \% & $1.33\pm 0.24\pm 0.05$ &
$1.32\pm 0.24\pm 0.05$ \\
$|\delta|$ & $(1.57\pm 0.15 \pm 0.05) \cdot 10^{-3}$ & 
$(1.57\pm 0.15 \pm 0.05) \cdot 10^{-3}$ \\
$ \mathrm{arg} \, \delta$ & $12.6^0 \pm 3.7^{\circ}\pm 0.2^{\circ}$ &
$13.0^{\circ} \pm 3.7^{\circ} \pm 0.2^{\circ}$ \\
$\beta$ (GS) & $-0.0695\pm 0.0053$ & --- \\
$a$ (HLS) & --- & $2.336\pm 0.016\pm 0.007$ \\ 
$\chi^2/\nu$ & 0.92 & 0.94 \\ \hline
\end{tabular}
\end{table}

In Table \ref{rhocom} the final results obtained in the GS model are 
compared to the world average  values \cite{PDG}.  
The values of the  $\rho(770)$ mass and width shown there are based
on the previous measurements at VEPP-2M
\cite{OLYACMD}, ALEPH \cite{ALEPH} and CLEO \cite{CLEO}. While the
mass of the $\rho(770)$ obtained in this work is in very good 
agreement with the previous measurements, our value of the
width is 2.7 standard deviations lower. This difference
can partly be explained by the difference of our parameterization
compared to the previous works. Particularly,
the value of $\Gamma_\rho$ is correlated with 
the value of $\mathrm{arg}\, \delta$. In
previous papers the parameter $\delta$  was assumed
to be real. Fixing $\mathrm{arg}\, \delta = 0$ increases our value 
of $\Gamma_\rho$ by 2 MeV.
Such model uncertainties as the effect of the complex phase of $\delta$
and the difference between the results of GS and HLS fits were not
included into the systematic error of the fit parameters.

The leptonic width of the $\rho(770)$ is
in good agreement with the result of \cite{OLYACMD} quoted by
\cite{PDG}. Our value of the branching ratio
$\omega \to \pi^+\pi^-$
is 1.6 standard deviations lower than the world average $(2.1\pm 0.4)\%$
based on the two most precise measurements from $e^+e^-$
experiments \cite{OLYACMD,DM1}. Again, a different parameterization
of the form factor was
used in previous works. Also note that the parameters of the
$\omega(782)$-meson, such as mass, width and the leptonic width, have
changed, that also affects the extracted value of the branching
ratio. Our fit to the old data \cite{OLYACMD} gives the value 
$Br(\omega\rightarrow\pi^+\pi^-)=(2.00 \pm 0.34)\%$,
still 1.6 standard deviations above our present result.

\begin{table}
\caption{\label{rhocom}Comparison of the fit to $|F_\pi(s)|^2$ in the GS
model to the world average values}
\vspace{1mm}
\begin{tabular}{lcc}
\hline
Parameter & GS model & World Average \\ \hline
$\mrho$, MeV & $776.09\pm 0.64\pm 0.50$ & $775.7 \pm 0.7$ \\ 
$\Gamma_\rho$, MeV & $144.46\pm 1.33 \pm 0.80$ & $150.4 \pm 1.6$ \\
$\Gamma(\rho\rightarrow e^+e^-)$, keV & $6.86\pm 0.11\pm 0.05$ &
$6.77 \pm 0.32$ \\ 
$Br(\omega\rightarrow\pi^+\pi^-)$, \% & $1.33\pm 0.24\pm 0.05$ &
$2.1 \pm 0.4$ \\
\hline
\end{tabular}
\end{table}

\subsection{Hadronic contribution to the muon anomalous magnetic moment} 

Let us estimate the implication of our results for 
$a_{\mu}^{\pi\pi}$,  the 
contribution from the annihilation into two pions, which dominates
the hadronic contribution to (g-2)/2. 
To this end we compare its value in the energy range studied in
this work and calculated from CMD-2 data only 
to that  based on the 
previous $e^+e^-$ measurements  \cite{OLYACMD,DM1}. Table \ref{g-2} 
presents results of the $a_{\mu}^{\pi\pi}$ calculations performed using
formula \eqref{dint} and the 
direct integration of the experimental data over the energy 
range studied in this work. As was explained
above, the bare cross section $\sigma^{0}_{\pi\pi(\gamma)}$ 
was used in the calculation. 
The method is straightforward and has been described
elsewhere \cite{EJ}. 
The first line of the Table (Old data) gives the result based 
on the data of OLYA, CMD and DM1 while the second one (New data) 
is obtained from the CMD-2 data only. 
The third line (Old + New) presents the 
weighted average of   
these two estimates. The assumption about the 
complete independence of the old and new data used in the
averaging procedure seems to be well justified. For 
convenience, we list separately statistical and systematic uncertainties
in the second column while the third one gives the total error 
obtained by adding them in quadrature. One can see that the 
estimate based on the CMD-2 data is in good agreement with that 
coming from the old data. It is worth noting that
the statistical error of our measurement is slightly larger 
than  the systematic uncertainty. 
Because of the small systematic error of the new data, 
the uncertainty of the new result for $a_{\mu}^{\pi\pi}$ is 
almost three times better than the previous one. 
As a result, the combined value based on both old and new data is 
completely dominated by the CMD-2 measurement. 

The example above only illustrates the importance of the improved 
accuracy. At the present time the analysis of the $\pi\pi$ data 
as well as other hadronic final states
in the whole energy range accessible to CMD-2 is in progress. 
Independent information is also available \cite{BES} or expected in close
future from other experiments studying
low energy $e^+e^-$ annihilation \cite{ISR}. 
When all the above mentioned data are
taken into account, one can expect
a significant improvement of the overall error  of $a_{\mu}^{had}$
(by a factor of about 2) compared to the previous one 
based on the $e^+e^-$ data only \cite{EJ}.   
 
\begin{table}
\caption{\label{g-2}Contributions of the $\pi\pi$ channel 
to (g-2)/2}
\begin{tabular}{lcc}
\hline
Data  & a$_{\mu}^{\pi\pi}$, 10$^{-10}$ & Total error, 10$^{-10}$ \\
\hline
Old   & 374.8 $\pm$ 4.1 $\pm$ 8.5 & 9.4 \\
\hline
New & 368.1 $\pm$ 2.6  $\pm$ 2.2 & 3.4 \\
\hline
Old + New & 368.9 $\pm$ 2.2 $\pm$ 2.3 & 3.2 \\ 
\hline
\end{tabular}
\end{table}    

The completion of the analysis of the $\pi\pi$ data will also 
provide a possibility of the precise test of the CVC based relation
between the cross section of $e^+e^- \to \pi^+\pi^-$ and the
spectral function in the decay $\tau^- \to \pi^-\pi^0\nu_{\tau}$.
The solution of the problem of the possible deviation between
$e^+e^-$ and $\tau$-lepton data \cite{CLEO,e20} should also involve 
a thorough investigation of the effects of isospin breaking
corrections as well as additional radiative corrections in 
$\tau$ decays \cite{kirill,ku,cir}.

\section{Conclusion}

The following values of the $\rho$ and $\omega$ meson parameters have
been obtained with the Gounaris-Sakurai fit to the formfactor data:
\begin{equation*}
\begin{array}{ll}
\mrho  & = (776.09 \pm 0.64 \pm 0.50)~ \text{MeV},
\vphantom{\Bigl( \Bigr)} \\
\Gamma_\rho  & = (144.46\pm 1.33 \pm 0.80)~ \text{MeV},
\vphantom{\Bigl( \Bigr)} \\
\Gamma(\rho\rightarrow e^+e^-)  & = (6.86\pm 0.11 \pm 0.05)~ \text{keV},
\vphantom{\Bigl( \Bigr)} \\
Br(\omega\rightarrow\pi^+\pi^-) & = ( 1.33\pm 0.24 \pm 0.05) \% ,
\vphantom{\Bigl( \Bigr)} \\
\mathrm{arg} \, \delta & = 12.6^{\circ} \pm 3.7^{\circ} \pm 0.2^{\circ}. 
\vphantom{\Bigl( \Bigr)} 
\end{array}
\end{equation*}

The measurement presented in this paper supersedes the preliminary
result \cite{INP9910}, obtained from the same data.
Analysis of the much larger data sample collected by CMD-2 in
1996 (the 370-540 MeV energy range), 1997 (the 1040-1380 MeV energy
range) and 1998 (the second scan of the 370-960 MeV energy range)
is in progress. 

\section{Acknowledgements}

The authors are grateful to the staff of VEPP-2M for excellent 
performance of the collider, to all engineers and technicians 
who contributed to the CMD-2 experiment. We acknowledge 
numerous useful discussions with M. Benayoun, A. H\"{o}cker, 
F. Jegerlehner, W.J. Marciano, K.V. Melnikov and G.N. Shestakov. 

This work is supported in part by grants DOE DEFG0291ER40646,  
INTAS 96-0624, Integration A0100, NSF PHY-9722600, NSF PHY-0100468
and RFBR-98-02-17851.

\end{document}